\documentclass[runningheads]{llncs}
\pdfoutput=1
\usepackage{graphicx}
\usepackage{url}
\usepackage{booktabs}
\usepackage[utf8]{inputenc}
\begin{document}
\title{Designing a Maturity Model for a Distributed Software Organization. An Experience Report\thanks{This work is part of a project that has received funding from the European Union's Horizon 2020 research and innovation programme under Grant Agreement No. 856726 (GN4-3).\protect\\ The scientific/academic work is financed from financial resources for science in the years 2019-2022 granted for the realization of the international project co-financed by Polish Ministry of Science and Higher Education.
}}
\author{Bartosz Walter\inst{1} \and
Marcin Wolski\inst{1} \and \\
\v{Z}arko Stanisavljevi\'{c}\inst{2} \and
Andrijana Todosijevi\'{c}\inst{3}
}
\authorrunning{B. Walter, M. Wolski et al.}
\titlerunning{Constructing a Maturity Model for a Distributed Software Organization}
\institute{
Pozna\'n Supercomputing and Networking Center, Pozna\'n, Poland \\
\email{\{bartek.walter, marcin.wolski\}@man.poznan.pl} 
\and
School of Electrical Engineering, University of Belgrade, Belgrade, Serbia
\\
\email{zarko.stanisavljevic@etf.bg.ac.rs}
\and
AMRES, Belgrade, Serbia
\\
\email{andrijana.todosijevic@amres.ac.rs}
}

\maketitle              % typeset the header of the contribution
\begin{abstract}
We present early experiences with defining and validating a software maturity model (SMM) for a distributed, research-driven organization of independent and self-organizing teams of diverse cultures,  experience and maturity. The paper briefly outlines the model, but focuses on the early stages of building and validating it. Based on that, we identify major factors contributing to the successful deployment of a SMM.

\end{abstract}

\section{Introduction}
\label{sec:intro}
Maturity of software teams reflects their ability to effectively address and implement all objectives, requirements and commitments related to constructing software. Software Process Improvement (SPI) is a general approach that encompasses the processes and practices involved in software development. It is generally accepted that software processes need to be continuously evaluated and optimized in order to better fulfill the expectations of the stakeholders of software projects~\cite{fuggetta2014software}. 

Maturity models (MMs) identify the objectives relevant for dedicated application domains, and measure the performance of subject organizations with respect to these objectives. MMs usually serve three primary goals: (i) identify the key elements that help to successfully deliver software, (ii) mark out the improvement directions for organizations, and (iii) provide a method of evaluating the maturity. Although several MMs are advertised as generic, they are actually domain-bound or implicitly include various contextual issues and constraints unique to a given organization. A systematic literature review (SLR) revealed that most of the published maturity models are based on practices and success factors from individual projects that showed good results in specific organizations or industries, although they lack a sound theoretical basis and methodology~\cite{Garcia-Mireles2012}. 
There are two main trends in software maturity evaluation. One of them concentrates on confronting the existing practices with a generic, stable model, which identifies the recommended elements of the process. Such models mandate the recognition and implementation of certain activities defined by the model, and are commonly used in practice, e.g., CMMI~\cite{Paulk1993} or TMM~\cite{Burnstein1996-tmm}.
The other approach refers to common values or principles (e.g., agility or the use of Scrum), such as Progressive Outcomes~\cite{Fontana2015}, that could be freely operationalized by the subject organizations. It leaves more freedom to the organizations that can easily adapt the framework to their needs and specifics. However, regardless of the chosen approach, the implementation of a MM may trigger organizational handicaps, resistance or even open opposition from the subject teams and units. That is why the process of deploying the model also deserves a thorough consideration and may substantially affect the effects.

In our previous paper~\cite{Stanisavljevic2018} we presented a preliminary version of a maturity model for G\'{E}ANT, a research-driven, distributed organization providing networking capabilities for researchers and software-bound services built on top of the pan-European network. Such a model would enable better harmonization of development processes and could promote exchange of good practices. In this paper we focus on observations from early evaluation of the model in a selected sample of software development teams. %In principle, we followed a multiple-step approach~\cite{Bruin2005-mma} with some modifications.
Based on them, we formulate recommendations concerning the process of deploying the maturity model. %The findings may be interesting for practitioners, as they provide practical guidelines for improving the adoption of MM within organizations, but also for researchers, as a source of empirical insights into a deployment process of a maturity model.

The paper is organized as follows. In Sec.~\ref{sec:relwork} we briefly outline the state-of-the art for maturity models, and existing works on the process of their definition and deployment. Next, in Sec.~\ref{sec:background} we describe context, i.e., present  G\'{E}ANT and its specifics. Sec.~\ref{sec:process} presents the process of defining and validating the model, considering also the communication with stakeholders. Finally, in Sec.~\ref{sec:lessons} we formulate lessons and recommendations extracted from the observations, and in Sec.~\ref{sec:conc} we provide concluding remarks.

\section{Related Work}
\label{sec:relwork}

%we should add here some stats - e.g. how many MM has been developed and proposed, 
%below there is a copy-paste from the cited article - it must be REPHREASED
% \cite{Garcia-Mireles2012}  many proposals of maturity models in the relevant literature.. Mettler and Rohner [3] indicate that they found a list of 135 different maturity models related to the discipline of information systems, while de Bruin et al. [1] found more than 150 models of maturity that assess, for instance, the maturity of IT service capability, strategic alignment of innovation management, program management, knowledge management and enterprise architecture.
%In the field of software engineering, von Wangenheim et al. [8] identified 52 software process capability/maturity models (SPCMM).

%not sure what should i do with the paragraph above, this is a copy from where, this Garcia cite? 

A variety of maturity models for various activities within software development have been developed and are in use. An SLR by Garc{\'{i}}a-Mireles, Moraga and Garc{\'{i}}~\cite{Garcia-Mireles2012} reports 35 different maturity models related to the discipline of information systems, more than 150 models that assess, e.g., the maturity of IT service capability, strategic alignment of innovation management, program management, knowledge management and enterprise architecture, as well as more than 50 software process capability/maturity models (SPCMM). 

In order to create the IT Performance Measurement Maturity Model (ITPM), Becker et al. compared several different maturity models which had a well-documented development processes~\cite{becker2009developing}. The first model chosen is the Capability Maturity Model Integration (CMMI), being a widely accepted method of evaluating the capabilities of software vendors, as a refinement of CMM~\cite{Paulk1993}. 
Analysis Capability Maturity Model (ACMM) is another model, in which some of the elements are adopted from the CMMI, but it is more related to analysis and operations research~\cite{covey2005creation}. The third one is Business Process Management Maturity (BPMM) model, representing a method for evaluation of BPM capabilities and achievements~\cite{Bruin2007}.% Last but not least, Document Process Maturity Model (DPMM) is also considered~\cite{Visconti1993}.

%Kulkarni and Freeze~\cite{kulkarni2004development} were the first to deal with the assessment and estimation of knowledge management capability (KMCA) in different knowledge areas. 

Furthermore, the process of information and communication technology (ICT) evolving was followed by development of the ICT management capability maturity framework~\cite{renken2004developing}. Iterative modeling, in combination with interviews with experts, made the framework a valuable tool for assessing the management capability of an organisation.     

%** Testing~\cite{burnstein1996developing}  ???

%Perhaps we can divide the literature somehow, into two groups 
%a) general MM models, CMMI-DEV, ITIL, 
Improvement paths in software engineering are defined by the guidelines given by CMMI-DEV and the international standard ISO/IEC 15504~\cite{von2010systematic}.
Although CMMI is considered a rather formal approach, the concept of maturity has been also transferred to other, less restricted environments~\cite{Stanisavljevic2018}. Open-source communities developed a number of models that reflect the capability of the projects, teams and individuals, e.g., Capgemini's Open Source Maturity Model (C-OSMM)~\cite{duijnhouwer2003open}, Qualification and Selection of Open Source (QSOS)~\cite{origin2006method} or OMM for open-source communities~\cite{Petrinja2009}. 
In~\cite{picard2015maturity}, the authors presented an ITIL-based maturity model that guides organizations on their way to an ISO/IEC 20000-1 certification and discusses its adequacy for evaluating the ISO/IEC 20000 certification readiness. 

% tailored MM, like for agile teams
Moreover, some studies consider maturity in terms of conformance to agility. Agile Maturity Model (AMM)~\cite{Patel2009} defines levels of agility, which increasingly address the common agile practices and values starting from basic ones, e.g., planning and requirements management, up to managing uncertainty and defect prevention. 

%** SLR on SMM
%~\cite{von2010systematic}  up

%MW - addressing reviewers comments
With respect to the process of constructing a maturity model, Bruin et al. pointed out that it should follow defined development phases~\cite{Bruin2005-mma} and outlined such a process. Also a work by P{\"o}ppelbu{\ss} et al.~\cite{Pppelbu2011-WhatMA} attempted to identify generic design principles for designing a maturity model. The authors of~\cite{munoz2016method} showed a method that provided strategies for the implementation of software process improvements based on the contextual aspects in which the software is developed. Finally, in~\cite{fontana2018maturity}, the authors looked at combinations of CMMI-DEV with agile methods. Based on the 14 models found, they analyzed levels’ structures, maturity concept, assessment methods and mature practices.

%\url{https://aisel.aisnet.org/cgi/viewcontent.cgi?referer=&httpsredir=1&article=1220&context=acis2005}
%~\cite{Bruin2005-mma}

%Telfor GN SMM for Software
%~\cite{Stanisavljevic2018}

%bibliografy - good source for other materials
%\url{https://wiki.geant.org/display/gn43wp9/References+and+related+work}

%IT Performance Measurement Maturity Model (ITPM)    %up
%https://link.springer.com/article/10.1007/s12599-009-0044-5
%~\cite{Becker2009}  

%SLR on SMM definition process~\cite{Garcia-Mireles2012}   %up

%\emph{Review Paper 3}
%https://ieeexplore.ieee.org/abstract/document/5728832/
%\cite{Unterkalmsteiner2012}
%In this paper the authors did a systematic literature review which included 148 papers published between 1991 and 2008 concerned with software process improvement. This paper represented our basis for defining measurable parameters in our instrument for conducting maturity assessment.
%MW: I added to the process part - its important paper and our bibliography lacked it

\section{Background and Context}
\label{sec:background}

% on the impact of contextual features of the project/organization on the optimization efforts
% about geant - standard description; highlight its unique features
G\'{E}ANT is a research and innovation organization, built upon a federation of NRENs -- operators of national networks for science and education. The NREN community is involved in collaborative software development activities, focused on delivering software products that provide advanced services\footnote{\url{https://www.GEANT.net/Resources/Media_Library/Documents/services_brochure_web.pdf}}. As a result, a number of software products have been developed and are currently in use (see Table~\ref{tab:geant}). They include projects of diverse maturity, size and target domains: starting from prototype (proof of concept) solutions, through pilot applications for a closed group of users, to mature products, supporting the delivery of services in production.% Their users mostly include G\'{E}ANT partners and their member institutions, researchers, students and educators, who expect high-quality, reliable services and infrastructure to support their work or studies. 

Most of software teams in G\'{E}ANT (SW teams) are distributed and involve engineers from different NRENs and the G\'{E}ANT organization. The teams are autonomous in adopting specific methodology for software development and choosing the assisting tools~\cite{Bilicki2014}. %Software developers (SW devs) from NRENs commit to the project objectives, project governance principles and adherence to the Project Lifecycle Management (PLM) outline.
That results in diversity of the processes and approaches to software development.
Most of the teams are small and prefer an agile (iterative) approach to software development. However, the level of adoption of best practices varies, and teams have struggled with their consistent adoption~\cite{Wolski2017}. 

% BW: outline the specific constraints in GN
Based on a comprehensive analysis, a need for optimization and coordination of the software development process has been identified~\cite{Wolski2017}. %That would require defining a model, targeted at identifying the current status and providing guidance for optimization. This process should embrace both all relevant perspective: SW development, service operations and management. 
To address this need, the Software Maturity Model (SMM)~\cite{Stanisavljevic2018} was designed, to provide a reference framework for adopting and customizing software development processes in an efficient and effective way, along with a method for evaluating the performance of SW teams and providing them with recommendations for improvements.

%Hereby, it should be pointed out that, besides the considerations related to a diverse G\'{E}ANT environment, the initiation of the SMM creation, and the process of its defining later on, actively involved SW teams and people with enough experience and expertise, requesting for a change and helping to define its direction. In order to fulfill the organization vision and objectives, the identified need for optimization and coordination of the software development process has been addressed, placing it in alignment with the SPI Manifesto [???].

In particular, SMM is expected to address the following objectives: 
\begin{itemize}
%BW: discuss - we need to focus on defining SWD objectives, and identifiyng the best practices; other goals and less prominent
\item to identify processes and activities essential for effective software development,
\item to help the teams in evaluating and optimizing their processes,
\item to promote and coordinate the dissemination of best practices among the teams.
%\item to act as a framework for the evaluation of SW teams' maturity and help in elaborating recommendations for improvements, 
%\item build the connection between the software teams and service delivery processes,
%\item support the achievement of better service maturity models. 
\end{itemize}

%Expected benefits that come from having an unified SMM include, but are not limited to:

%\begin{itemize}
%\item definition of concise set of software best practices, that can improve the work process and effectiveness of  SW teams (small, medium and large teams),
%\item delivery of a flexible tool dedicated for SW teams to retrospect their processes in order to identify the areas for improvement,
%\item more matured SW teams in terms of improved and more effective software processes aligned with other activities, and supported with validated solutions.
%\end{itemize}

%it fits to the conclusion
%The G\'{E}ANT project has the central coordination and management entity, which provides governance in the area of IPR, policy for managing the software artifacts (code, documentation, database), and verifies the compliance with Product Lifecycle Management (PLM). The MM validation has been coordinated and supervised by this entity.  

\subsection{ G\'{E}ANT software projects}
\label{ssec:projects}

Software projects could be categorized with respect to common features that define their context. For example,~\cite{Capiluppi2003} identified 12 such features (project age, application domain or programming language) to present the characteristic of open source projects. 
In Table~\ref{tab:geant} we present selected information for G\'{E}ANT software projects:
\begin{itemize}
\item Codebase age -- time between the first and last commit in the project,
\item Codebase size -- SLOC,
\item Languages -- number of languages used in a project,
\item Team size -- number of contributors for the project in all roles (tester, developer, manager etc.),  
\item Team size (software developers) -- number of software developers contributing to the project,
\item Projects per person -- a number of projects a single person contributes to.
\end{itemize}

\begin{table}[]
\centering
\caption{G\'{E}ANT software projects and teams as of April, 2019}
\label{tab:geant}
\begin{tabular}{|l|l|l|l|l|}
\hline
Statistic name & \# items & Min & Max & Median \\ \hline
\textbf{Software projects} & 26 &  &  &  \\ 
Codebase age & & 11 days & 17 years & 3 years \\ 
Codebase size & & 1058 & 506531 & 15893 \\ 
Languages & & 1 & 23 & 9 \\ \hline
\textbf{Software teams} & 24 &  &  &  \\
Team size & & 1 & 11 & 3 \\ 
Team size (Software developers) & & 1 & 6 & 2 \\
Project per person & & 1 & 4 & 1 \\ \hline
\textbf{Software contributors} & 168 &  &  &  \\
\textbf{Software developers} & 34 & &  &  \\ \hline
\end{tabular}
\end{table}

%Currently, there are 26 projects in the  G\'{E}ANT software portfolio. They are usually developed by the G\'{E}ANT and NREN community, but in few cases they have been inherited from external open source initiatives, and are currently developed collaboratively with them (e.g., COTURN\footnote{\url{ https://github.com/coturn/coturn/blob/master/README.md}}). Some characteristics of the software development in G\'{E}ANT, like partial and scattered involvement in specific projects, are common for the FOSS development model (free and open-source software)~\cite{Dangle2005}   

%There are 24 software teams with at least one person on board who is a present member of the G\'{E}ANT project. One team is solely dedicated to one software project. Moreover, there are 168 G\'{E}ANT software contributors, who either commit source codes or report issues. 34 of 168 G\'{E}ANT contributors are software developers, who have produced at least 1000 LOC. %(internal definition of a software developer, to distinguish between software developers and other members who have made the commits only occasionally).
%The G\'{E}ANT product portfoliocontains the projects with different size (from 1058 up to 506 531 LOC), system age (from 11 days up to more than 17 years). Usually a software project is developed using a number of different languages (9 languages on average).
%Noteably, G\'{E}ANT software teams are usually small: they include 3 members on average, 2 of them being software developers. The largest team has 11 members with 6 developers. In most cases, every person contributes to only one project, but there are exceptions.

\subsection{Outline of the Maturity Model}
\label{ssec:outline}
\begin{figure}
  \caption{Outline of the SMM design process}
  \label{fig:model}
  \includegraphics[width=\textwidth]{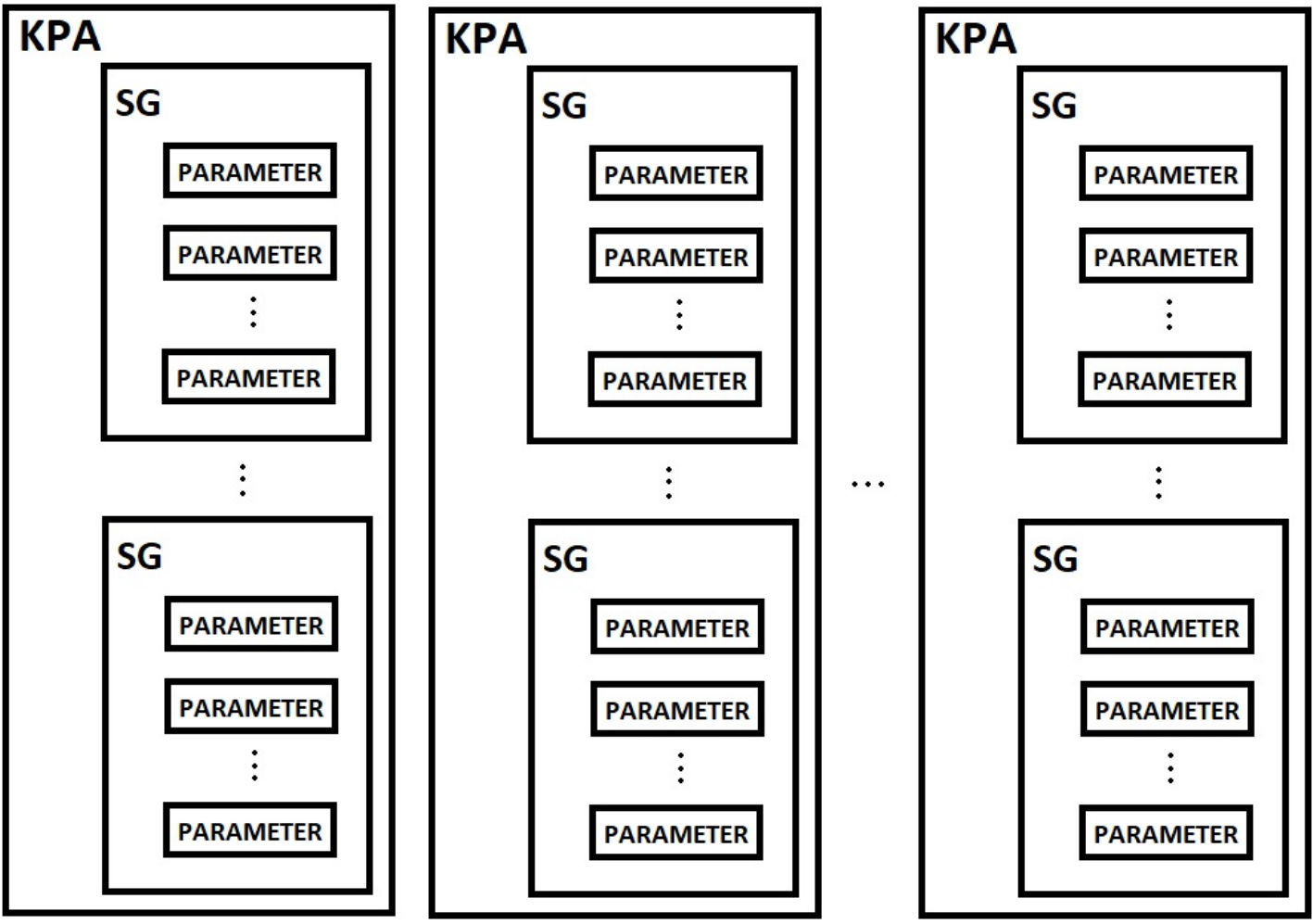}
\end{figure}

The preliminary version of the G\'{E}ANT MM~\cite{Stanisavljevic2018} is founded on the structure of CMMI. It comprises of five Key Performance Areas (KPAs): Requirements Engineering, Design and Implementation, Quality Assurance, Team Organization and Software Maintenance. Specifically, we introduced the latter two KPAs to address two distinctive issues G\'{E}ANT is facing, concerning distributed teams and continual maintenance of the software products. Each KPA defines a number of Specific Goals (SGs), describing objectives that should be addressed by the teams. To facilitate their evaluation, we adopted a set of parameters that capture various dimensions of the goals. Finally, values on each parameter were added to produce an aggregate quantitative assessment of each KPA. Interpretation of the KPA score combined with results of individual SGs provided the input for producing feedback of recommended improvements for the teams.
The general model structure is depicted in Fig.~\ref{fig:model}, and more details can be found in ~\cite{Stanisavljevic2018}. However, there is an ongoing work on adjusting the model and making its elements more precise, so its structure and scope are subject to evolve.

\section{Process of Defining the Maturity Model}
\label{sec:process}

\begin{figure}
  \caption{Outline of the SMM design process}
  \label{fig:process}
  \includegraphics[width=\textwidth]{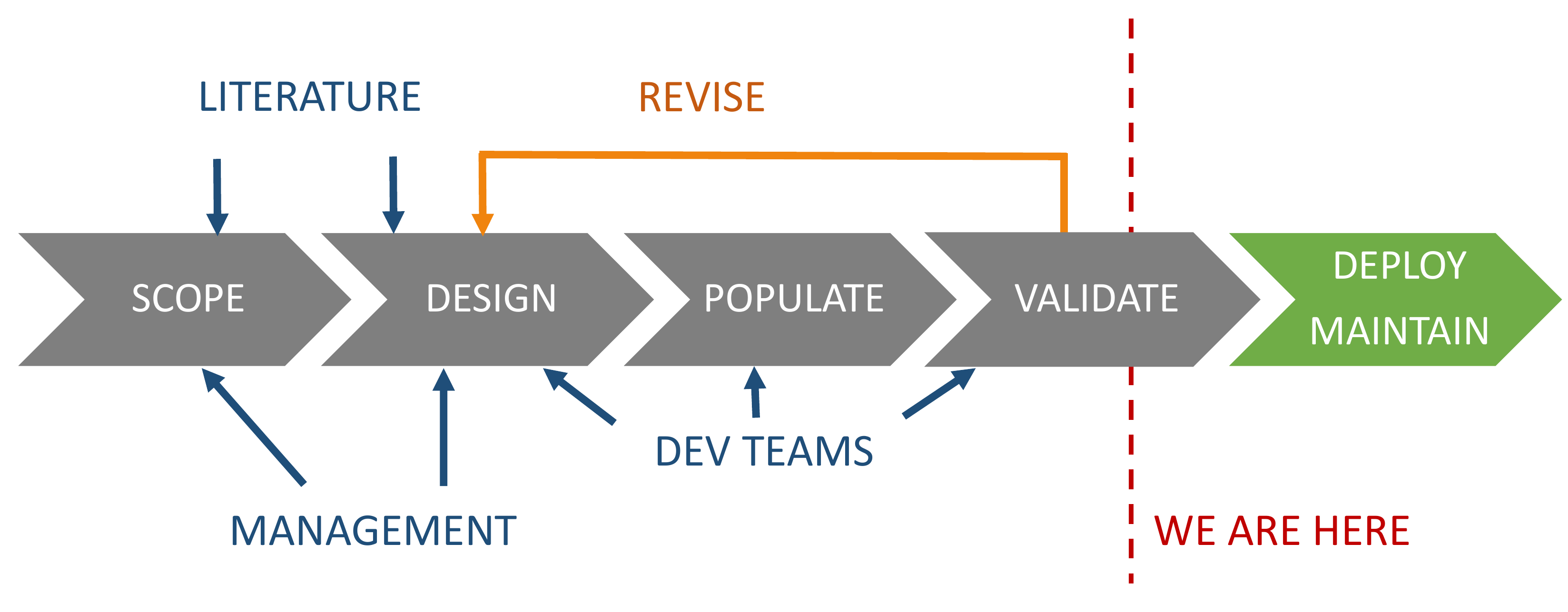}
\end{figure}

The process of defining, validating and deploying a maturity model has a significant impact on the final result. In this section we outline the process we followed, inspired by recommendations in~\cite{Bruin2005-mma,Pppelbu2011-WhatMA}. They are presented in Fig.~\ref{fig:process} and discussed in the following sections.

%HERE a diagram presenting the MM development process should be added

% recommendations in literature concerning the types of MMs, and the approach to defining them
% MOVED to sec. relwork

%We decided to design our maturity model, based on Review Paper 1 and Review Paper 2, and for the population of parameters we used Review Paper 3.

% I'd suggest taking the article: Understanding main phases of developing MAM and  refer to the figure 1: model development phases: scope, design, populate, test, deploy, maintain. Describe in the corresponding sections each phases:
%Scope - 
%Design - 
%....

%the creation of an assessment questionnaire - can we add a point here? 
%I mentioned it in the description of review paper 

\textbf{Scope --}
In this phase we identify the expectations, scope and future stakeholders of the desired model. The primary goal is to define the boundaries for model application. 

We started with a literature review focused on identification of suitable existing models and methods of constructing new models. Ramasubbu et al.~\cite{Ramasubbu2005} noticed that CMM lacks KPAs  addressing the capabilities for managing distributed software projects and they identified 24 new KPAs for that purpose. A work by Dangle et al.~\cite{Dangle2005} provided insights into constructing custom maturity models atop CMM and adjusting them to specific needs. The paper~\cite{Unterkalmsteiner2012} represented our basis for defining measurable parameters in our instrument for conducting maturity assessment. Finally, the importance of maintenance-related activities, outlined in~\cite{April2009}, inspired us to address this area in the MM for G\'{E}ANT.

Maturity models have either descriptive or prescriptive purpose, i.e., the intention to describe the as-is-situation or to define a road map for improvements~\cite{Stanisavljevic2018}. The literature shows that the maturity model design principles can be organized in that manner~\cite{Pppelbu2011-WhatMA}. We decided to focus on prescriptive purpose, so that SMM could be used to recommend improvements in the development process.

Initially, we invited two types of stakeholders: management and development teams. Later, using the snowballing technique, we also included DevOp engineers, responsible for operating the software systems after they are deployed. 

After a discussion with stakeholders, we identified the primary objective for the SMM: creating a mechanism for assisting and guiding the teams in improving their processes, based on empirical data, considering the G\'{E}ANT structure and specific constraints. Perspective of evaluating the teams appeared not so important, as it sparkled doubts and worries among development teams. 

The model identifies \em{process areas} and \em{specific goals} that the subject SW teams should address, while the teams can choose the exact methods of doing that, given the fact that the framework gives a basis for objective evaluation of teams’ performance, indicating directions for improvements.

\textbf{Design --}
This phase is expected to determine the architecture for the proposed model. We started with performing an extensive literature research, aimed at extracting different types of maturity models. 
%MW: here maybe references to relwork section? or give a few examples from there?
The analysis of the existing MMs helped us gain a comprehensive understanding of the real situation in software development and to better define scope and design of our framework.  

We proposed a hierarchical, three-level structure of areas, goals and parameters, essential for effective software development.
For our model we seek for a balance: the model adopts the structure of CMMI as a core, but also incorporates several goals and activities specific to agile development methods, which would produce a model tailored to the needs of innovation-driven, distributed organization~\cite{Stanisavljevic2018}.
Since the structure involves areas not present in other maturity models that are specific to G\'{E}ANT, we frequently consulted the designed elements of the model with selected representatives of all groups of stakeholders. 

Furthermore, we chose a continuous representation for evaluating the maturity. It enables measuring the team’s maturity in different areas separately, without producing a single aggregate score, and helps the teams to focus on the most relevant processes. This approach is also more flexible than staged representation, since it allows accommodating changes in the future more easily ~\cite{Stanisavljevic2018}.

\textbf{Populate --}
This phase is aimed at identification of the specific goals, as well as the methods of evaluating them.

This task needs to be performed in a close collaboration with subject teams, as they can explicitly indicate activities essential for the final successful deployment of the model. We started by selecting the goals from CMMI and adding some other, based on G\'{E}ANT specifics. Next, they were iteratively discussed and analyzed with selected representatives of the development teams.
An important, but underestimated element of defining the goals is to supplement them with appropriate description and make their language layer uniform, to reduce possible misunderstanding and confusion.

For evaluation of the goals, we initially proposed a number of common parameters, representing the terms and concepts that are related with each goal. Each parameter was to be evaluated in a 3-level scale: missing, addressed internally (within the subject team) and addressed externally (by an external authority, e.g., G\'{E}ANT managers, professional standards or the NRENs). Later, the evaluation scheme was changed to a 5-level one, in response to remarks submitted by the development teams.

\textbf{Validate --}
Once the model is populated, it needs to be tested for relevance and rigor. Both the structure and content, as well as the created model instruments need to be validated.

We performed early validation conducting evaluation with 3 selected teams (see Table~\ref{tab:validation}), with an objective to validate the model rather than the teams. The invited teams were diverse with respect to their size, distribution, and type of developed products.
%MW I don't know why but there is 4 on the PDF (instead of table.2)? FIXED

\begin{table}[]
\caption{G\'{E}ANT teams taking part in early validation of the SMM}
\label{tab:validation}
\begin{tabular}{|l|l|l|r|r|}
\hline
\toprule
Team Id& Domain & Product Age & Involved NRENs & Team size \\ \hline
\midrule
A & Network Management & 2 years 2 mons & 4 & 5 \\ \hline
B & Network Connectivity & 5 years 8 mons & 6 & 11 \\ \hline
C & Trust and Identity & 4 years 1 mon & 4 & 4 \\ \hline
\end{tabular}
\end{table}

The evaluation took the form of a direct or remote interview with a member of the subject development team, guided by a member of the SMM team. Subjects were asked to describe how they address the subsequent goals and comment on inconsistencies, missing or redundant element of the model. Their responses were recorded in questionnaires by the SMM member, and further analyzed. To reduce team-specific bias, two members of the subject development team were interviewed separately, and the differences were discussed.

This phase resulted in identification of several issues in the model and the interview questionnaire. They led to re-designing the model and the questionnaire, and another phase of the validation. After the pilot evaluation is performed, the framework was adjusted and changes based on the pilot evaluation were incorporated.

\textbf{Deploy and Maintain --}
At this stage, the model is ready for deployment and use in the G\'{E}ANT environment.  
These phases have not been reached yet.

\section{Lessons Learned}
\label{sec:lessons}

Our experiences in defining the SMM in G\'{E}ANT currently embrace the pilot phase, from inception to early validation. Although it does not cover the entire model life-cycle, we can draw some early conclusions and recommendations, discussed below, based on observations we made. We need to highlight that the findings are related only to the process, and not to the subject organization.

\medskip
\noindent
\emph{Lesson 1: Identify and involve relevant stakeholders}

A maturity evaluation usually involves various stakeholders with their unique perspective on the goals and objectives. It is highly important to early identify all stakeholders, both the ones that could contribute to the model and those affected by it. A failure in achieving that may result in two effects: (i) constructing an incomplete model, that does not capture all relevant perspectives, or (ii) fomenting conflicts among the stakeholders that may hinder or even block the maturity improvement efforts. 

Within a software organization, the maturity evaluation comprises both managerial and technical domains. Involving senior managers in this effort and acquiring support from them can substantially facilitate the process of constructing the model.

\medskip
\noindent
\emph{Lesson 2: Identify and clearly present the objectives}

% removing the preliminary comparison
% practice-orientation - OK
% identification of needs - oK
% the expectation for over-specification or over-generalization - OK
% reconciliation of different needs of the stakeholders - TODO

The perspectives of various stakeholders may differ or even conflict. Therefore, the objectives defined for the maturity model should be explicitly declared, then shared and agreed upon by all stakeholders. That minimizes the risk of hidden intentions that could void the effort invested in constructing the model.

Regardless of the stated objectives, a maturity model always involves an element of evaluation. Development teams can consider it as an instrument that could be used against them, which could put them in an opposition to the improvement efforts. To mitigate the risk, the process of defining and deploying the model should be transparent and integral. In particular, the communication addressing various stakeholders should be consistent. If the model is updated, the changes should be disseminated to all parties as soon as possible, to prevent misinformation and confusion leading to tensions.

\medskip
\noindent
\emph{Lesson 3: Capture contextual factors}

Maturity discussed in literature is frequently referring to an idealistic model that does not address specifics of individual organizations. In case of minor differences, they could be easily addressed and reconciled within generic models, but some factors play too important role to be ignored. In response to that, custom models seek to identify the uniqueness of activities in the subject organization.

We recommend addressing this issue two-fold: major contextual items should be incorporated to the model, as they shape the specific \emph{modus operandi}, unique to the organization, while the minor ones could be approached during evaluation. For example, geographical distribution significantly changes the way how the organization operates, and should be addressed by the model, while the use of a specific toolset is not a major factor in attaining the goals defined in the model.
The typical contextual factors include structure of organization, level of customer involvement, method of communication or required level of accountability.

\medskip
\noindent
\emph{Lesson 4: Early and frequently validate the model with stakeholders}

In initial stages of defining the model, the changes are frequent and extensive, which may confuse stakeholders. To prevent that, the model should be frequently validated, e.g., by running pilot evaluations, which additionally highlight the role of the stakeholders in the process. That gives both parties, the stakeholders and the SMM team, a chance to identify discrepancies and early resolve possible issues.

The feedback plays a central role in the process. Each raised issue should be identified, recorded, discussed, tracked and resolved, to promote transparency of the process. Otherwise, a feedback debt is likely to occur, which accumulates negative emotions and deprives the model from the positive improvements it could bring.

If new stakeholders are identified during the SMM development, they should be included into discussion as soon as possible, but after being on-boarded by the SMM team. That prevents other stakeholders from loosing the focus while the new members catch up with the progress.

\medskip
\noindent
\emph{Lesson 5: Refine the model iteratively}

The feedback and new ideas should be incorporated to the model in proper batches. Revising the model in response to every single issue may result in a never-ending loop of tiny improvements, which hinder keeping the focus and tracking the changes. As a side effect, discrepancies between subsequent versions are inevitable and are likely to grow.
On the other hand, a policy of large steps may also yield a sense of confusion and lack of continuity between the concepts highlighted in various versions. Then, we found a lightweight, incremental approach with time-boxed release periods for publishing subsequent revisions of the model to work best for us.

%\medskip
%\noindent
%\emph{Lesson 6: Collect data and inform??} 
% aggregation and presentation of results
% method of data collection - questionnaire

\section{Conclusions}
\label{sec:conc}

% where we are now, what has been achieved
% how well the goals have been accomplished
% what could be improved
% next steps

%Bartek+Andrijana

Typically, software organizations apply maturity models not only for assessing and improving their processes, but also for monitoring their optimization efforts. They can be particularly useful for distributed organizations that share a common culture, but apply different practices, like G\'{E}ANT. It also addresses the requirements of SPI Manifesto\footnote{\url{http://2019.eurospi.net/images/eurospi/spi_manifesto.pdf}}, by acknowledging the combination of people and processes as a crucial factor of effective SPI projects~\cite{Korsaa2012}.

However, it is crucial to set the solid grounds and clear basis of how the SMM framework should be developed and composed. It is essential to get the comprehensive understanding of the whole community and relevant stakeholders in order to set and define the contextual factors that are specific to the organization. Moreover, evaluation and continuous improvement in early stages of the framework development are more than beneficial and represent one of the most important phases. We found iterative approach to be the best path in the process of making the model transparent and creating the positive attitude towards its deployment. 

So far, we have received support from the G\'{E}ANT software community and gained valuable feedback from selected representatives, who really helped us to better understand the requirements and working patterns of G\'{E}ANT teams.

We are currently at the end of the pilot evaluation phase. The next step would be a second revision of the SMM, based on the collected results and knowledge, as well as incorporation of the agreed changes. The ultimate goal is to provide guidance and a set of best practices for G\'{E}ANT software teams and community, which would help in improving the software development process.

\bibliographystyle{unsrt}
\bibliography{smm-definition}

\end{document}